\documentclass[10pt,conference]{IEEEtran}
\usepackage{cite}
\usepackage{enumerate}
\usepackage[inline]{enumitem}
\usepackage{amsmath,amssymb,amsfonts}
\usepackage{algorithmic}
\usepackage{graphicx}
\usepackage{textcomp}
\usepackage{xcolor}
\usepackage{url}
\usepackage[a4paper,hmargin=13mm,vmargin={25mm,30mm},headsep=4mm]{geometry}
\def\BibTeX{{\rm B\kern-.05em{\sc i\kern-.025em b}\kern-.08em
   T\kern-.1667em\lower.7ex\hbox{E}\kern-.125emX}}

\begin{document}
\title{VOSySmonitoRV: a mixed-criticality solution on Linux-capable RISC-V platforms}

\author{Flavia Caforio, Pierpaolo Iannicelli,  Michele Paolino,  Daniel Raho \\
Virtual Open Systems,\\
17 Rue Lakanal, 38000 Grenoble, France\\
\texttt{\{f.caforio, p.iannicelli, m.paolino, d.raho\}@virtualopensystems.com}\\
\texttt{\url{http://www.virtualopensystems.com}}
}

\maketitle

\begin{abstract}
Embedded systems are pervasively used in many fields nowadays. In mixed-criticality environments (automotive, industry 4.0, drones, etc.) they need to run real-time applications with certain time and safety constraints alongside a rich operating system (OS). 
This is usually possible thanks to virtualization techniques, that leverage on hardware virtualization extensions on the machine. However, these hardware extensions might not cope with the security and safety requirements of the specific use case, and additionally, they might not always be available. A notable example is the emerging RISC-V architecture, that is today gaining a lot of traction in the mixed criticality field, but that does not offer today hardware virtualization extensions.
In this paper VOSySmonitoRV is proposed as a mixed-criticality solution for RISC-V systems. 
VOSySmonitoRV allows the co-execution of two or more operating systems in a secure and isolated manner by running in the highest privileged machine level.
A specific benchmark, measuring the interrupt latency and context switch time is done to assess the system performance in mixed criticality systems.
\end{abstract}


\section{Introduction}
In the new era of Industry 4.0 \cite{simo2021role}, the importance of solutions to allow the co-execution of activities with different levels of criticality in common hardware platforms has become crucial.
It is now a sought-after that in recent embedded systems there is the need to have at the same time features typical of generic operating systems (for connectivity, user interface, etc.) together with the ones of a real time operating systems (for interaction with critical sensors, rotors, motors, etc.).
The most common solution to run multiple operating systems in a single platform is the use of virtualization. However, standard hypervisor solutions are designed and implemented focusing on the performance of the guests. This makes difficult to address the safety and security requirements of mixed criticality systems.

In this context, the RISC-V architecture \cite{WatermanI} is quickly gaining relevance in the industry and is bringing a new set of challenges and solutions regarding both safety and security. The use of standard virtualization solutions to implement mixed criticality solution is not possible today. In fact, instruction set extensions to support hypervisors have not yet reached the standard specification, and today there is no hardware platform in the market than can use them.

This paper proposes VOSySmonitoRV, a mixed critical solution for RISC-V platforms that does not need instruction set extensions to support hypervisors. VOSySmonitoRV is able to run multiple operating systems in a single platform partitioning the system resources and protecting the execution of real time/safety critical operating systems.
Performance measurements related to interrupts and context switch operations, considered as key points for multi-operating systems performance, are presented to demonstrate the feasibility of such a design on RISC-V platform.
This paper is organized as follows: Section 2 details the main features of VOSySmonitoRV, Section 3 presents first prototype benchmarks done on the interrupt and context switch latency, Section 4 describes the related work and Section 5 concludes the paper. 



\section{VOSySmonitoRV}\label{vosys}
VOSySmonitoRV is a mixed-criticality solution that provides spatial and temporal isolation between co-executing operating systems (OSes) on a multi-core RISC-V processor. It is a software layer that handles the memory area partitioning and isolation in terms of memory, permissions, peripherals, interrupts, and harts (i.e., CPU cores).
In fact, VOSySmonitoRV makes sure that the safety-critical and trusted OSes are booted before un-trusted OSes, guaranteeing for the former best boot time and certifiability. In addition, VOSySmonitoRV is able to provide additional services to the operating systems running on the platform, e.g., power management, communication between OSes, trusted execution environment, custom vendor specific functions, etc.

\begin{figure}[htbp]
\centerline{\includegraphics[width=0.8\columnwidth]{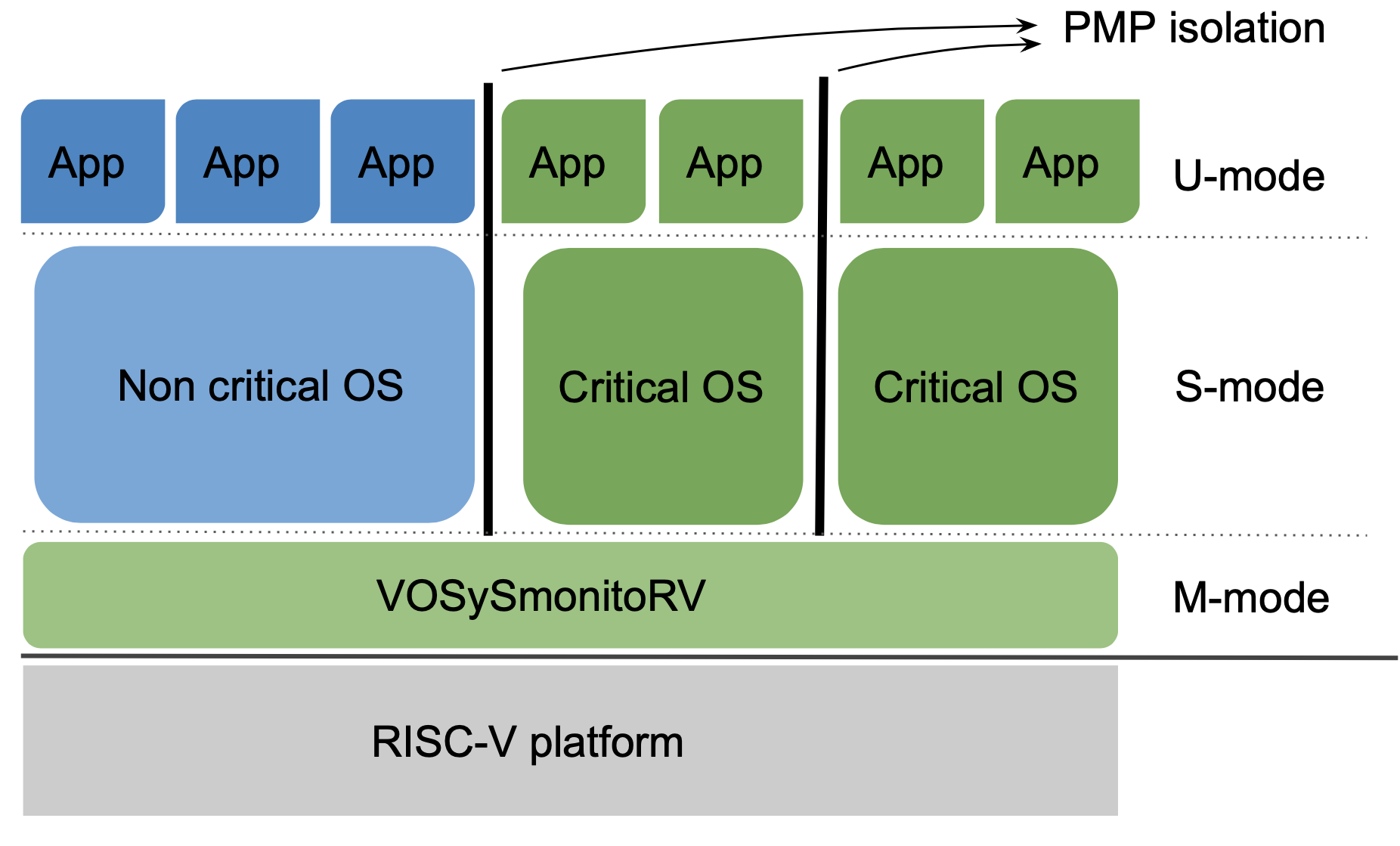}}
\caption{VOSYSmonitoRV architecture overview}
\label{vosys_ov}
\end{figure}

As shown in Figure \ref{vosys_ov}, VOSySmonitoRV exploits standard features of RISC-V: 
\begin{enumerate*}
\item it leverages on the RISC-V's privileged architecture, implementing all the VOSySmonitoRV software running in M-Mode;
\item it partitions the memory using the Physical Memory Protection (PMP) unit from the RISC-V privileged ISA standard  \cite{Waterman}, defining memory areas (and their permissions) that are used to run the different operating systems.
\end{enumerate*}
VOSySmonitoRV main benefits are security and isolation provided by technologies like PMP, toghether with the possibility to be already installed in existing RISC-V platforms with no hypervisor extensions,

\subsection{First VOSySmonitoRV prototype}
The first VOSySmonitoRV prototype has been implemented to concurrently FreeRTOS as real time/safety critical operating system and Linux v5.8 as generic and feature rich operating system. 
At the system boot time VOSySmonitoRV is executed and initializes the system, setting up low level drivers and taking care of the system resources partitioning for FreeRTOS and Linux. More in particular, the PMP device is configured to guarantee isolation between operating systems, isolating their memory, devices and harts.

\begin{figure}[htbp]
\centerline{\includegraphics[width=1\columnwidth]{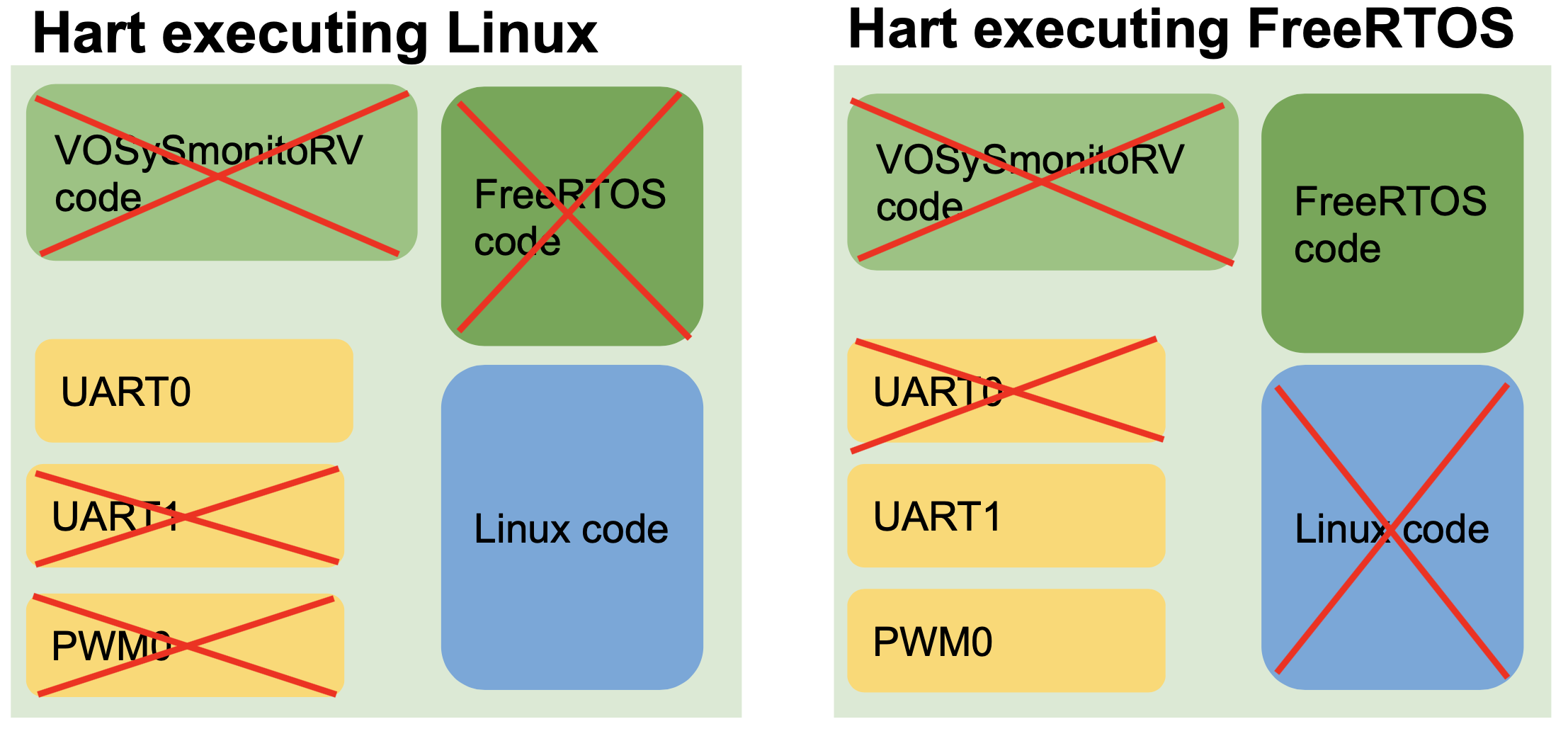}}
\caption{Memory permissions to U-mode and S-mode during different OS execution in the first prototype of VOSySmonitoRV}
\label{pmpfig}
\end{figure}

As shown in Figure \ref{pmpfig}, each operating system has its own PMP configuration that details access control to memory and devices. Both Linux and FreeRTOS are not able to read/modify the VOSySmonitoRV nor the other operating system memory, thus protecting the integrity of the system.
Serial devices (UART0 and UART1) are allocated respectively to Linux and FreeRTOS, isolating the two operating systems console input/output. This approach can be applied to any other memory mapped devices in the system.
VOSySmonitoRV is executed in M-mode of all processors and can access to all memory regions that are not locked. This places VOSySmonitoRV in a good position to monitor the execution of the operating systems running in the board, and potentially reboot one of them to react to specific safety/security events occur. 

The Linux operating system does not need any specific modification to be un on VOSySmonitoRV. 
For that concerns FreeRTOS, minimal modifications were needed to enable execution in S-mode. In fact, the currently available open source version of FreeRTOS automatically runs in M-mode. This might be acceptable in low-power devices that are not equipped with with M, S and U execution modes. However, in a context of a multicore platform we consider a better choice to run FreeRTOS is S-mode, which is the standard execution mode for operating system kernels (e.g., Linux).
The FreeRTOS extensions developed to support S-mode execution include: i) register access adaptation to use the ones accessible in S-mode instead of the ones accessible in M-mode, ii) set up and configuration of interrupt handling related registers and iii) configuration of timer settings and CPU release via ECALLs (the privileged instruction of the RISC-V privileged ISA \cite{Waterman}). 

The VOSySmonitoRV first prototype, statically allocating one hart per operating system has been implemented using the SiFive HiFive Unleashed, a development platform for SiFive's FU540-C000 SoC \cite{sifivefu540}.
A video demonstration has been realized \cite{vosysdemo} to show VOSySmonitoRV in action.


\begin{figure*}[htbp]
\centerline{\includegraphics[width=2\columnwidth]{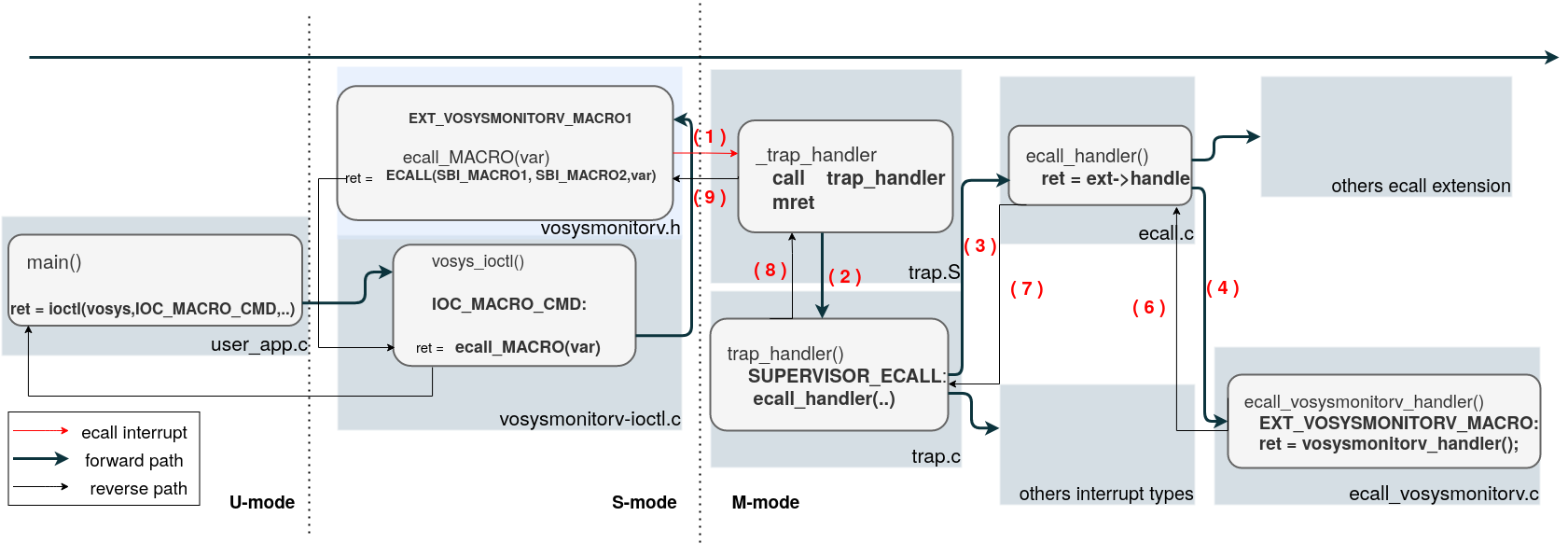}}
\caption{ECALL interrupt and context switch benchmarks flow diagram with interaction steps brackets}
\label{flow}
\end{figure*}

\section{VOSySmonitoRV benchmark}
The VOSySmonitoRV first prototype performance has been measured to assess its overhead and responsiveness in mixed critical applications, as well as to assess the feasibility of the described approach.
More in particular, the ECALL interrupt latency and the context switch overhead from Linux to M-mode has been measured to assess the time needed by an operating system to request one of the services provided by VOSySmonitoRV or to trigger a context switch between operating systems when the multiple OSes hart sharing function will be implemented.
To measure this, the cost in terms of clock cycles to handle an ECALL has been measured on the aforementioned SiFive board.

\subsection{Environment and implementation details}
For the implementation of this benchmark the following components were developed: i) a Linux application (U-mode), ii) a Linux kernel driver (S-mode) and iii) a dedicated benchmark extension in VOSySmonitoRV (M-mode). The Linux application makes an ioctl call to request the execution of specific ECALLs by the Linux driver.
These ECALLs trigger the execution of a purpose-built handler in VOSySmonitoRV. 
The Linux driver and the purpose built benchmark extension in VOSySmonitoRV are the two endpoints where a performance counter is started/stopped to gather performance metrics. 
Figure \ref{flow} shows the interaction between these components and the steps executed by the different actors.

Benchmark measurements are focused mainly to assess ECALL interrupt and context switch performance. More in details:
\begin{itemize}
\item \textbf{ECALL interrupt}: 
\begin{itemize}
\item \textbf{Latency} from the driver to the ECALL VOSySmonitoRV handler function, or in other words, the delay between the ECALL occurring and the first instruction in the handler of that interrupt. This is the path from step \texttt{(1)} to step \texttt{(4)} in Fig.\ref{flow}.
\item  \textbf{Overhead}: from the driver to the ECALL VOSySmonitoRV handler function and back. This is the path from step \texttt{(1)} to step \texttt{(9)} in Fig.\ref{flow}. It represents a full timing overhead to handle a very simple ECALL. 
\end{itemize}
\item \textbf{Interrupt context switch}: 
\begin{itemize}
\item \textbf{Save context}: from the driver to the VOSySmonitoRV main trap handler. It mainly consider the hardware latency and the saving of the context (32 registers). This is the transition done in step \texttt{(1)} (Fig.\ref{flow}).
\item \textbf{Restore context}: from the VOSySmonitoRV handler to the driver. It consider the restore of the context (32 registers) and the hardware latency. This is the transition of step \texttt{(9)} (Fig.\ref{flow}).
\end{itemize}
\end{itemize}

Performance counter is read using the \texttt{rdcycle} pseudo instruction at each of the two endpoints detailed in the steps above and computing the difference between the two values.
For example in the \textbf{Interrupt context switch} (Save context) measurement, the performance counter is read in the driver before the ECALL and then again after the context saving. 

All the measurements are done 500 times each to collect the results and do statistical consideration about the values. 
Moreover, as suggested by the official iRISC-V documentation \cite{WatermanI}, Linux is forced to run with one processor core (kernel boot argument \textit{maxcpus = 1}) to overcome existing hardware issues to isolate the number of clock cycles per hart.
 
\subsection{Results}
Figure \ref{context} below shows the \textbf{ECALL interrupt latency} and \textbf{ECALL interrupt handling overhead} results, showing an average latency time of 0.46 $\mu$S and a complete ECALL handling time of 0.66 $\mu$S.
These results are prone to software performance variations due to caches and interrupts, resulting in variable values and a standard deviation of 0.066$\mu$S in the former and 0.076$\mu$S in the latter case. 

\begin{figure}[htbp]
\centerline{\includegraphics[width=0.8\columnwidth]{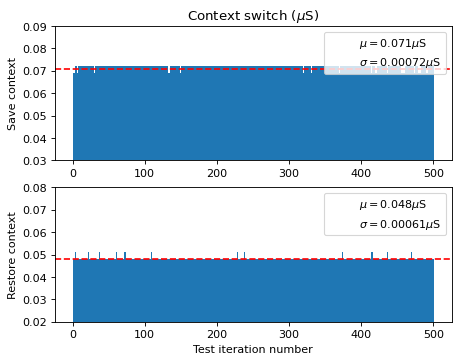}}
	\caption{Context switch measurements, with an average value ($\mu$) between 0.07 and 0.4$\mu$S and a standard deviation ($\sigma$) close to 0$\mu$S. }
\label{mincontext}
\end{figure}

Conversely, results shown in Fig.\ref{mincontext} for the \textbf{Interrupt context switch} present a lower standard deviation, mainly because there is little/no software variance that impacts this test. Context saving operations need about 0.071 $\mu$S while restore operation cost about 0.048 $\mu$S. 

\begin{figure}[htbp]
\centerline{\includegraphics[width=0.8\columnwidth]{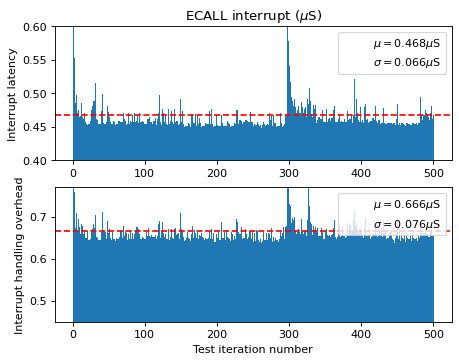}}
	\caption{Interrupt latency measurements, with an average value ($\mu$) between 0.4 and 0.6$\mu$S and a standard deviation ($\sigma$) in the range of 0,06 and 0.07$\mu$S. }
\label{context}
\end{figure}

The results of these benchmarks are encouraging if we compare them with similar benchmarks done in the past by the authors. In the Arm case, for instance, measured context switch of the ASIL-C certified VOSySmonitor product resulted in value in the range between 0.6 $\mu$S to 1.4$\mu$S \cite{lucas2017vosysmonitor}.

The comparison between different architectures is out of the scope of this work. However, by proving better performance than an ASIL certified product on Arm architecture, VOSySmonitoRV demonstrates to benefit by an architecture and implementation that can cope with mixed critical requirements. Lastly, this work also confirms the readiness of RISC-V processors to execute multiple operating systems in a mixed criticality environment.


\section{Related works}
VOSySmonitoRV combines strong isolation typical of Trusted Execution Environments (TEE) with functionalities of a hypervisor, aimimng directly from the design phase to a certifiable mixed critical solution for automotive and industrial applications.

For what concerns TEEs, key existing solutions are today MultiZone
\cite{garlaticlean} and KeyStone \cite{lee2020keystone}. Both solution, similarly to VOSySmonitoRV, do not require specific hardware extensions to work. 
MultiZone features include a real time scheduler and a formally verifiable code base. KeyStone, on the other hand has the possibility to run FreeRTOS in one of its enclaves. 
However, both their focus is mainly on providing security for application, in some cases specifically built for the purpose of being executed in an enclave.
More recently, SiFive presented WorldGuard, a security hardware model that offers World ID markers to every hart and every process to protect and isolate different domain execution (data and code).
Being orthogonal to VOSySmonitoRV, such solution can be integrated in a VOSySmonitoRV to provide stronger isolation.
An other security related example is \textbf{Donky} \cite{schrammel2020donky}, a solution that is a hardware-software co-design for memory isolation of user processes. It has a secure monitor called Donky Monitor, part of the software design, that handles in-process access policies in user space. Donky is used for memory isolation of protection keys and it has no kernel interaction, using the RISC-V 'N' extension (the user-level interrupts extension).

For what concerns virtualization and hypervisors, as mentioned previously the standard specification does not yet include virtualization extensions and there is no real hardware implementing them.
Hypervisors such as Bao \cite{martins2020bao} are developed using the QEMU emulator and cannot guarantee the performance and flexibility given by hardware support for virtualization. In general these solutions are prototyped waiting for the availability of hypervisor extensions in RISC-V hardware.
VOSySmonitoRV solution has the advantage is to enable co-existance of multiple operating systems without virtualization extension that could introduce large attack surface due to implementation complexity \cite{sierra2020security}. 


\section{Conclusion}
This paper proposes VOSySmonitoRV, a mixed critical solution for multicore RISC-V platforms. VOSySmonitoRV uses the standard RISC-V instruction set architecture does not need virtualization extensions to run multiple operating systems in a single platform. 
This is very important in an evolving architecture like RISC-V, where virtualization extensions are stil missing in hardware platforms available the market.
Performance measurements are presented, showing an \textbf{ECALL interrupt latency}  time value of 0.46 $\mu$S and a \textbf{Interrupt context switch} equal to 0.071 $\mu$S.
Interrupt latency and context switch are considered key operations for such as system, because these are an indicator of the time needed by VOSySmonitoRV to switch operating system or to full fill a request coming from an application (e.g., power management).
With such numbers, VOSySmonitoRV demonstrates to benefit by an architecture and implementation that can cope with mixed critical requirements, as well as it proves the readiness of RISC-V processors to execute multiple operating systems in a mixed criticality environment.
Future works include the development of the hart-sharing feaure, to enable multiple operating systems to share hart resources in a configurable way and respecting security and safety requirements of mixed criticality systems.

\section*{Acknowledgment}
This project has received funding from the EU Horizon 2020 Programme under grant agreement No 957269 (EVEREST).


\bibliographystyle{./IEEE_bibitex/IEEEtran}
\bibliography{./conference_101719}

\begin{thebibliography}{10}
\providecommand{\url}[1]{#1}
\csname url@samestyle\endcsname
\providecommand{\newblock}{\relax}
\providecommand{\bibinfo}[2]{#2}
\providecommand{\BIBentrySTDinterwordspacing}{\spaceskip=0pt\relax}
\providecommand{\BIBentryALTinterwordstretchfactor}{4}
\providecommand{\BIBentryALTinterwordspacing}{\spaceskip=\fontdimen2\font plus
\BIBentryALTinterwordstretchfactor\fontdimen3\font minus
  \fontdimen4\font\relax}
\providecommand{\BIBforeignlanguage}[2]{{%
\expandafter\ifx\csname l@#1\endcsname\relax
\typeout{** WARNING: IEEEtran.bst: No hyphenation pattern has been}%
\typeout{** loaded for the language `#1'. Using the pattern for}%
\typeout{** the default language instead.}%
\else
\language=\csname l@#1\endcsname
\fi
#2}}
\providecommand{\BIBdecl}{\relax}
\BIBdecl

\bibitem{simo2021role}
J.~Sim{\'o}, P.~Balbastre, J.~F. Blanes, J.-L. Poza-Luj{\'a}n, and A.~Guasque,
  ``The role of mixed criticality technology in industry 4.0,''
  \emph{Electronics}, vol.~10, no.~3, p. 226, 2021.

\bibitem{WatermanI}
A.~Waterman and K.~Asanovi\'{c}, \emph{“The RISC-V Instruction Set Manual,
  Volume I: User-Level ISA, Document Version 2019121}, December 2019.

\bibitem{Waterman}
\emph{The RISC-V Instruction Set Manual Volume II: Privileged Architecture
  Version 1.10}, May 2017.

\bibitem{sifivefu540}
\emph{\BIBforeignlanguage{English}{SiFive FU540-C000 Manual}}, SiFive, Inc.

\bibitem{vosysdemo}
\BIBentryALTinterwordspacing
V.~O.~S. Inc. (2020) Vosysmonitorv: a mixed criticality virtualization solution
  for risc-v. [Online]. Available:
  \url{http://www.virtualopensystems.com/en/solutions/demos/vosysmonitorv-risc-v-demo/}
\BIBentrySTDinterwordspacing

\bibitem{lucas2017vosysmonitor}
P.~Lucas, K.~Chappuis, M.~Paolino, N.~Dagieu, and D.~Raho, ``Vosysmonitor, a
  low latency monitor layer for mixed-criticality systems on armv8-a,'' in
  \emph{29th Euromicro Conference on Real-Time Systems (ECRTS 2017)}.\hskip 1em
  plus 0.5em minus 0.4em\relax Schloss Dagstuhl-Leibniz-Zentrum fuer
  Informatik, 2017.

\bibitem{garlaticlean}
C.~Garlati and S.~Pinto, ``A clean slate approach to linux security risc-v
  enclaves.''

\bibitem{lee2020keystone}
D.~Lee, D.~Kohlbrenner, S.~Shinde, K.~Asanovi{\'c}, and D.~Song, ``Keystone: An
  open framework for architecting trusted execution environments,'' in
  \emph{Proceedings of the Fifteenth European Conference on Computer Systems},
  2020, pp. 1--16.

\bibitem{schrammel2020donky}
D.~Schrammel, S.~Weiser, S.~Steinegger, M.~Schwarzl, M.~Schwarz, S.~Mangard,
  and D.~Gruss, ``Donky: Domain keys--efficient in-process isolation for risc-v
  and x86,'' in \emph{29th $\{$USENIX$\}$ Security Symposium ($\{$USENIX$\}$
  Security 20)}, 2020, pp. 1677--1694.

\bibitem{martins2020bao}
J.~Martins, A.~Tavares, M.~Solieri, M.~Bertogna, and S.~Pinto, ``Bao: A
  lightweight static partitioning hypervisor for modern multi-core embedded
  systems,'' in \emph{Workshop on Next Generation Real-Time Embedded Systems
  (NG-RES 2020)}.\hskip 1em plus 0.5em minus 0.4em\relax Schloss
  Dagstuhl-Leibniz-Zentrum f{\"u}r Informatik, 2020.

\bibitem{sierra2020security}
F.~SIERRA-ARRIAGA, R.~BRANCO, and B.~LEE, ``Security issues and challenges for
  virtualization technologies,'' 2020.

\end{thebibliography}

\end{document}